\documentstyle[prb,multicol,aps,epsf]{revtex}
\begin{document}
\draft
\title{\bf Optical study of the metal-nonmetal 
transition in Ni$_{1-\delta}$S }
\author{H. Okamura} 
\address{Department of Physics, Faculty of Science, 
Kobe University, Kobe 657-8501, Japan.}
\author{J. Naitoh and T. Nanba} 
\address{Graduate School of Science and Technology, 
Kobe University, Kobe 657-8501, Japan.}
\author{M. Matoba, M. Nishioka, and S. Anzai} 
\address{Department of Applied Physics and Physico-Informatics, 
Faculty of Science and Technology \\ 
Keio University, Yokohama 223-8522, Japan.}

\date{\today}
\maketitle
\begin{abstract}
Optical reflectivity study has been made on the hexagonal 
(NiAs-type) Ni$_{1-\delta}$S in order to probe its electronic 
properties, in particular those associated 
with the metal-nonmetal transition in this compound.    
Samples with $\delta$=0.005 and 0.02 are studied, 
which have transition temperatures $T_t$=246~K and 161~K, 
respectively.  
A pronounced dip appears in the reflectivity spectra 
upon the transition, and the optical conductivity spectra 
show that the electronic structure below $T_t$ is similar 
to that of a carrier-doped semiconductor with an energy gap 
of $\sim$~0.15~eV.   
The optical spectra indicate that the gap becomes larger with 
decreasing temperature, and it becomes smaller 
as $\delta$ increases.    
It is also found that the overall spectrum including 
the violet region can be described based on a 
charge-transfer-type semiconductor, consistent with recent 
photoemission results.
\end{abstract}
\vspace{.5cm}
\pacs{Keywords: D. optical properties, D. phase transitions}

\begin{multicols}{2}
\narrowtext

The problem of the metal-nonmetal phase transition in the 
hexagonal NiS has 
been studied for three decades, but the transition 
mechanism is not completely understood yet.    
The high temperature (HT) phase above the transition 
temperature, $T_t \sim$~260~K, is a paramagnetic metal.  
Upon cooling through 
$T_t$, the resistivity increases suddenly by a factor of 
$\sim$~40, associated with slight increase in the lattice 
constants (0.3~\% in $a$ and 1~\% in $c$) and the appearance 
of an antiferromagnetic order.\cite{sparks,anzai,koehler}    
$T_t$ is lowered sharply with increasing Ni vacancies, and 
the transition disappears when the vacancy content exceeds 
$\sim$~4~\%.\cite{ohtani}    Similar behavior is observed also 
with an applied pressure, and the transition is not observed 
at pressures above $\sim$~2~GPa.\cite{mcwhan}   
These behaviors are summarized in the phase diagram of 
Fig.~1.   

The nature of the low-temperature (LT) phase below $T_t$ has 
been studied by many experiments.   The resistivity ($\rho$) 
increases only slightly with cooling, with an activation energy 
of several meV.\cite{ohtani}    In contrast, an optical study by 
Barker and Remeika\cite{barker} clearly showed the presence of 
an energy gap of about 0.15~eV.     Hall effect experiment 
by Ohtani\cite{ohtani} has shown that the majority 
carrier in the LT phase is the hole, and the density of holes 
is proportional to that of Ni vacancies, with $\sim$ 2 holes 
per Ni vacancy.     Namely, the LT 
phase can be described as a $p$-type degenerate 
semiconductor, where the Ni 
vacancies act as acceptors.     
Effects of substituting other elements for Ni or S have 
been also studied in detail.\cite{matoba1,matoba2}      
Recently, two high-resolution photoemission 
studies\cite{nakamura,sarma} have revealed a finite 
density of states (DOS) around the 
Fermi energy ($E_F$) in the LT phase, but they have given 
contrasting interpretations: Nakamura {\it et al.}\cite{nakamura} 
have concluded that there is a small correlation-induced band gap 
with an unusually sharp band edge, and that the observed finite 
DOS at $E_F$ is due to thermal and instrumental broadenings of 
the edge.    
On the other hand, Sarma {\it et al.}\cite{sarma} have concluded 
that the LT phase is an ``anomalous metal''. 

Various models have been proposed to account for the phase 
transition and the gap opening in NiS.     
At early stage, it was proposed that the transition was a Mott-type 
transition.\cite{white}  Namely, the Ni 3$d$ band splits 
into two bands separated by a gap in the LT phase 
due to strong Coulomb interaction at Ni sites.    
It was also proposed that the semiconducting electronic 
structure was caused by the antiferromagnetic 
order.\cite{mattheiss}    
Band calculations that take into account many-body 
effects via various approximations have been unable to 
reproduce an energy gap with a reasonable magnitude.\cite{nakamura}  
More recently, 
it has been proposed, based on cluster-model calculations 
and photoemission experiments, that the energy gap in NiS 
is of charge-transfer type, namely the energy gap in NiS is 
formed between the upper 3$d$ and the 
S 3$p$ bands.\cite{fujimori1,fujimori2,fujimori3}

Here we report our optical 
reflectivity study on the electronic structure of Ni$_{1-\delta}$S.   
We observe large spectral changes upon the phase transition, 
showing the formation of an energy gap of $\sim$~0.15~eV 
in agreement with the earlier optical work.\cite{barker}  
Our present work, however, reveals 
more detailed temperature-dependece of the spectra, 
and provides a comparison between samples having different $T_t$.  
The optical gap in the LT phase becomes larger with 
decreasing temperature, and also the gap becomes 
smaller as $\delta$ increases.   In addition, we 
show that the overall spectra including the violet region 
can be understood assuming a charge-transfer semiconductor, 
consistent with the cluster-model 
calculations and photoemission 
experiments.\cite{fujimori1,fujimori2,fujimori3}

The samples used in this work were 
polycrystalline Ni$_{1-\delta}$S prepared as follows.   
Ni and S powders were mixed with mole ratios of 
[Ni]:[S]=1:1 and 0.98:1, and melted at 1000~C$^\circ$ 
in an evacuated quartz tube.    
Then they were annealed at 700~C$^\circ$ for 2 days and 
at 500~C$^\circ$ for 1 week, and quenched in iced water.   
The 1:1 mixture resulted in an ingot with $T_t$=246~K, 
and the 0.98:1 mixture with $T_t$=161~K.   
Comparing these $T_t$ values with a previously-reported 
$T_t~-~\delta$ diagram,\cite{matoba1} 
the vacancy contents in these samples are estimated to be 
$\delta~\sim 0.005$ (or Ni$_{0.995}$S, $T_t=246$~K) and 
$\delta~\sim 0.02$ (Ni$_{0.98}$S, $T_t=161$~K).\cite{footnote}   
The ingots were cut into disk-shaped samples, 
and the surface was mechanically polished with alumina powders.    
Then the samples were annealed at 500~C$^\circ$ for 3 days in an 
evacuated quartz tube, followed by a quench in iced water.   
This re-annealing process is necessary because the mechanical 
cutting and polishing suppress the phase transition in the 
sample surface.\cite{barker}   
Reflectivity measurements below 2.5~eV were done using a 
rapid-scan Fourier interfefometer (Bruker Inc. IFS-66v) 
and conventional sources.  
Measurements between 2~eV and 30~eV were done using 
synchrotron radiation source at the beamline 
BL7B of the UVSOR Facility, Institute for Molecular Science.  
Standard near-normal incidence configuration was used 
for the reflectivity measurements.   An optical conductivity 
spectrum $\sigma(\omega)$ was obtained from a measured 
reflectivity spectrum $R(\omega)$ using the Kramers-Kronig 
relations.\cite{wooten}    
Hagen-Rubens ($1-a \sqrt{\omega}$) 
and $\omega^{-4}$ exptrapolation functions were used to 
complete the lower- and higher-energy ends of the 
refelctivity spectra.  

Figure~2 shows the infrared $R(\omega)$ of Ni$_{0.995}$S 
and Ni$_{0.98}$S measured at several temperatures.   The inset shows 
the $R(\omega)$ of Ni$_{0.995}$S at room temperature up to 30~eV.    
For both samples, the most significant spectral change upon the 
transition is a large reduction of $R(\omega)$ in the 
infrared region, accompanied by a ``dip'' near 0.15~eV.   
The result for 
Ni$_{0.995}$S is very similar to the single crystal result of 
Barker and Remeika\cite{barker}.   
These spectral changes occurred over a temperature width 
of a few K around $T_t$, as observed on our rapid-scan 
spectrometer while cooling down slowly.    
With cooling further in the LT phase, 
the dip blue-shifts slightly, which is seen clearer for 
Ni$_{0.995}$S.   
Figure~3 shows the corresponding $\sigma(\omega)$ spectra.   
For both samples, $\sigma(\omega)$ has a sharp 
rise toward lower energy in the HT phase, which is typical 
of a good metal.   Below $T_t$, however, the spectral weight below 
$\sim$ 0.3~eV is strongly depleted with an ``onset'' of 
$\sigma(\omega)$ at $\sim$ 0.15~eV, which is indicated by 
the arrows in Fig.~3.  
Below the onset, there exists a small rise in $\sigma(\omega)$.   
This metallic, Drude-like spectral component indicates 
that there are a small but sizable amount of free carriers 
in the LT phase.  
These optical spectra suggest that the 
LT phase is a semiconductor with an energy gap of $\sim$ 0.15~eV and 
excess carriers, or equivalently a carrier-doped semiconductor.   
This is consistent with the Hall effect result,\cite{ohtani} 
which showed that there were rather high density of holes 
($\sim$ 10$^{21}$~cm$^{-3}$) in the LT phase.   Another possible 
interpretation for the LT phase from these optical spectra is a semimetal, 
that has a low carrier concentration and a strong reduction in 
the DOS around $E_F$, or a ``psedogap''.    
It is seen that the blue-shift of the dip in $R(\omega)$ with 
cooling has a corresponding 
blue-shift of the onset in $\sigma(\omega)$, 
which shows that the energy gap becomes larger with decreasing 
temperature.   
This behavior of NiS was previously unknown.    
Note that a measurement of $\rho(T)$ would not give this 
information, since for a $p$-type semiconductor the activation 
energy given by $\rho(T)$ is connected to acceptor-related 
states at low temperatures,\cite{seeger} rather than that 
associated with the intrinsic energy gap.   

Although the spectral changes upon the phase transition 
are qualitatively similar for the two samples, there are 
important differences: (i) The Drude-like 
component in the LT phase has much larger spectral weight 
for Ni$_{0.98}$S 
than for Ni$_{0.995}$S. (ii) The onset in $\sigma(\omega)$ is lower 
in energy for Ni$_{0.98}$S than for Ni$_{0.995}$S.     
(i) indicates that there are much more free carriers 
in the LT phase of Ni$_{0.98}$S than that of Ni$_{0.995}$S.   
This is consistent, since Ni$_{0.98}$S contains more 
holes than Ni$_{0.995}$S due to larger density of Ni 
vacancies.\cite{ohtani}    
(ii) indicates that Ni$_{0.98}$S has a smaller 
energy gap than NiS.    This observation implies that the 
energy gap in the LT phase becomes smaller as $\delta$ 
(Ni vacancy content) increases.    This is another important 
result in this work.   The reason for this gap narrowing 
with increasing $\delta$ is unclear at present.   
The lattice constants become smaller as $\delta$ 
increases,\cite{matoba1} which may broaden the bands 
above and below the gap, resulting in a narrower gap.  
On the other hand, increasing $\delta$ also leads to a 
larger carrier density and a larger acceptor-related DOS 
around $E_F$.   
It is likely that the observed gap narrowing and the 
lowering of $T_t$ with increasing $\delta$ result from 
a complicated interplay among these effects.  

Recently, Sarma {\it et al}\cite{sarma} have reported a 
high-resolution photoemission study of NiS.  They observed 
a metallic electronic structure in the LT 
phase, where the DOS at $E_F$ was nearly flat and smoothly 
varying, and was also slightly smaller than that in the HT phase.  
Contrasting the metallic electronic structure around $E_F$ 
to the weakly temperature-dependent 
$\rho$ and the large optical gap, they argued that 
the LT phase was an ``anomalous metal''.   
However, a metallic DOS {\it within the 
close vicinity of $E_F$} is not necessarily inconsistent with 
a small variation in $\rho(T)$ and a large gap in the LT phase, 
since these behaviors can be viewed as typical of a $p$-type, 
degenerate semiconductor.\cite{ohtani,seeger}   
Namely, in a degenerate $p$-type semiconductor, 
$E_F$ is located near the top of the valence band, 
where a large acceptor-related DOS is present.   
Then it is possible 
to have a metallic (continuous) DOS around $E_F$ which is smaller than 
that for the HT phase (good metal).  
For such case, the activation energy measured by $\rho(T)$ at 
low temperatures probes the activation of holes to these 
low-lying states near $E_F$, and it is not directly 
related to the intrinsic gap.   It has been demonstrated 
convincingly that the LT phase of NiS is a 
$p$-type, degenerate semiconductor by Ohtani\cite{ohtani}, 
and the present optical result gives further support 
to this picture.     

Although we have described the LT phase as a carrier-doped 
semiconductor, it does not mean at all that the LT phase is 
a ``conventional'' semiconductor.    The holes in the LT phase 
are probably under the influence of strong 
on-site Coulomb interaction, 
and it is likely that the holes do not show simple 
free-particle behaviors.   In this respect, the behaviors of 
the holes in the LT phase are very intersting and deserve 
further studies.   

Fujimori {\it et al}\cite{fujimori1,fujimori2,fujimori3} 
have performed cluster-model 
calculations for NiS that take into account many-body and 
inter-configuration interactions.   
Comparing the calculated results with the measured photoemission 
(PE) spectra, they have concluded that the energy gap in NiS is a 
charge-transfer (CT) gap that opens between Ni 3$d$- and 
S 3$p$-derived bands, rather than a Mott-Hubbard gap 
between the correlation-split 3$d$ bands.     
Figure 4 illustrates optical transitions in these 
two cases.   The gap excitation is a $p$-$d$ transition 
in the CT case, while it is a $d$-$d$ transition in the 
Mott-Hubbard case.   Note that a $d$-$d$ transition is optically 
forbidden, but vacancy-related disorder and $p$-$d$ 
hybridization may make the transition 
partially allowed.    Regarding the infrared gap of 
NiS as a CT gap, 
the broad absorption band in $\sigma(\omega)$ at 
0.2~eV~$\leq$ $\hbar \omega \leq$ 2.5~eV (Fig.~3) 
can be attributed to optical transitions from 
the S 3$p$ band\cite{footnote2} to the upper Ni 3$d$ band, or 
equivalently to charge transfers from S to Ni.   
Then, the weak peak near 4.5~eV in $\sigma(\omega)$, 
seen in the inset of Fig.~1, can be assigned to transitions 
from the lower 3$d$ to the upper 3$d$ bands (indicated by 
the dashed arrow in Fig.~4).    
The observed peak energy is in good agreement with the estimated 
on-site Coulomb interaction energy, $U = 4.0 \pm 0.5$~eV, 
given by Fujimori {\it et al}.\cite{fujimori2}

In conclusion, we have presented an optical reflectivity 
study of Ni$_{1-\delta}$S.     
The optical spectra in 
the infrared show large changes upon the phase transition, 
and the spectra for the LT phase can be described 
as a carrier-doped semiconductor with an energy 
gap of $\sim$ 0.15~eV.   The magnitude of the optical gap 
became larger with decreasing temperature, and it became 
smaller for larger $\delta$.    
The overall spectrum including the violet region can be 
understood based on a charge-transfer-type semiconductor, 
consistent with previous photoemission results and cluster-model 
calculations.    
Further work is in progress in wider ranges of 
$\delta (0.005 \leq \delta \leq 0.04$), temperatures 
(8~K~$\leq$ 295~K), and photon energies (down to far-infrared), and 
quantative analyses based on the sum rule for $\sigma(\omega)$ 
will provide further information on the metal-nonmetal 
transition of NiS.

We thank S. Kimura for providing the Kramers-Kronig analysis 
software used in this work.   We acknowlege financial support 
from Grants-in-Aid from the Ministry of Education, Science 
and Culture.

\begin{figure}
\caption{Schematic phase diagram of Ni$_{1-\delta}$S 
in terms of temperature ($T$), Ni vacancy concentration 
($\delta$), and external pressure ($P$). }
\end{figure}

\begin{figure}
\caption{(a) Infrared reflectivity ($R$) spectra of Ni$_{0.995}$S 
and Ni$_{0.98}$S measured at several temperatures.   
The inset shows $R$ of Ni$_{0.995}$S at 295~K up to 30~eV.}  
\end{figure}

\begin{figure}
\caption{Optical conductivity ($\sigma$) spectra 
of Ni$_{0.995}$S and Ni$_{0.98}$S below 2.5~eV at several 
temperatures.  The arrows indicate the ``onset'' discussed 
in the text.  The inset shows $\sigma$ of Ni$_{0.995}$S 
at 295~K up to 30~eV. }  
\end{figure} 

\begin{figure}
\caption{Illustration of optical excitations across the energy gap 
for a charge-transfer insulator (a) and a Mott insulator (b).   
In (a), the gap is formed between the Ni 3$d$ upper Hubbard 
band (UHB) and the S 3$p$ band, while in (b) it is formed 
between UHB and the lower Hubbard band (LHB).  }  
\end{figure}

\end{multicols}

\begin{references}
\bibitem{sparks} Sparks, J.T. \& Komoto, T., 
{\it Rev. Mod. Phys.} 
{\bf 40}, 1968, 752.  

\bibitem{anzai} Anzai, S \& Ozawa, K., 
{\it J. Phys. Soc. Jpn.} 
{\bf 24}, 1968, 271. 

\bibitem{koehler} Koehler, R.F. \& White, R.L. 
{\it J. Appl. Phys.} 
{\bf 44}, 1971, 943.  

\bibitem{ohtani} Ohtani, T. 
{\it J. Phys. Soc. Jpn.} 
{\bf 37}, 1974, 701. 

\bibitem{mcwhan} McWhan, D.B., Marezio, M., Remeika, J.P., \& Dernier, P.D. 
{\it Phys. Rev.} 
{\bf B 5}, 1972, 2552.

\bibitem{barker} Barker, A.S., \& Remeika, J.P. 
{\it Phys. Rev.} 
{\bf B 10}, 1974, 987. 

\bibitem{matoba1} Matoba, M., Anzai, S. \& Fujimori, A. 
{\it J. Phys. Soc. Jpn. }
{\bf 60}, 1991, 4230.  

\bibitem{matoba2} Matoba, M., Anzai, S. \& Fujimori, A. 
{\it J. Phys. Soc. Jpn. }
{\bf 63}, 1994, 1429.  

\bibitem{nakamura} Nakamura, M., Sekiyama, A., Namatame, H., 
Kino, H., Fujimori, A., Misu, A., Ikoma, H., Matoba, M., \& Anzai, S. 
{\it Phys. Rev. Lett.} 
{\bf 73}, 1994, 2891.  

\bibitem{sarma} Sarma, D.D., Krishnakumar, S.R., Chandrasekhara, N., 
Weschke, E., Sch\"{u}{\ss}ler-Langeheine, C., Kilian, L., \& Kaindl, G., 
{\it Phys. Rev. Lett. }
{\bf 80}, 1998, 1284.  

\bibitem{white} White, R.M. \& Mott, N.F. 
{\it Philos. Mag.} 
{\bf 24}, 1971, 845.  

\bibitem{mattheiss} Mattheiss, L.F. 
{\it Phys. Rev.} 
{\bf B 10}, 1974, 995.  

\bibitem{fujimori1} Fujimori, A., Matoba, M., Anzai, S. 
Terakura, K., Taniguchi, M., Ogawa, S., \& Suga, S. 
{\it J. Magn. Magn. Mater. }
{\bf 70}, 1987, 67.  

\bibitem{fujimori2} Fujimori, A., Terakura, K., 
Taniguchi, M., Ogawa, S., Suga, S., Matoba, M. \& Anzai, S. 
{\it Phys. Rev.} 
{\bf B 37}, 1988, 3109.   

\bibitem{fujimori3} Fujimori, A., Namatame, H., Matoba, M. \& Anzai, S. 
{\it Phys. Rev.} 
{\bf B 42}, 1990, 620.  

\bibitem{footnote}  It is possible that even carefully 
prepared sintered 
NiS with $T_t \simeq$ 264~K has a sizable amount of 
Ni vacancies, $\delta \leq$~0.002.  [See 
Sawa, T. \& Anzai, S. 
{\it J. Appl. Phys.} 
{\bf 49}, 1978, 5612.]  

\bibitem{wooten} Wooten, F. {\it Optical Properties 
of Solids} (Academic Press, New York, 1972).

\bibitem{seeger} See, for example, Seeger, K. 
{\it Semiconductor Physics: An introduction} 
(Springer Verlag, Berlin), Chapter 2. 

\bibitem{footnote2} Note that the ``3$p$ band'' here is not purely 
derived from S, but there should be a strong hybridization 
with the Ni 3$d$ band, since the gap is much smaller 
that the width of the absorption band.  


\end{references}
\end{document}